\documentclass[12pt]{iopart}
\usepackage{epsfig}
\usepackage{graphicx}
\usepackage{lscape}
\usepackage{longtable}
\usepackage{deluxetable}

\usepackage{iopams}  
\begin{document}

\title[Wide-orbit planets]{The Occurrence of Wide-Orbit Planets in Binary Star Systems}

\author{B. Zuckerman$^1$}

\address{$^1$Department of Physics and Astronomy, University of California, Los Angeles, CA 90095, USA}
\eads{\mailto{ben@astro.ucla.edu}}
\begin{abstract}
The occurrence of planets in binary star systems has been investigated via a variety of techniques that sample a wide range of semi-major axes, but with a preponderance of such results applicable to planets with semi-major axes less than a few AU.   We utilize a new method -- the presence or absence of heavy elements in the atmospheres of white dwarf stars -- to elucidate the frequency in main sequence binary star systems of planets with semi-major axes greater than a few AU.  We consider only binaries where a putative planetary system orbits one member (no circumbinary planets).  For main sequence binaries where the primary star is of spectral type A or F, data in the published literature suggests that the existence of a secondary star with a semi-major axis less than about 1000 AU suppresses the formation and/or long  term stability of an extended planetary system around the primary.  For these spectral types and initial semi-major axis $\geq$1000 AU, extended planetary systems appear to be as common around stars in binary systems as they are around single stars. 
\end{abstract}

$\smallskip$

{(stars:) binaries: general; (stars:) planetary systems; (stars:) white dwarfs}

\section{INTRODUCTION}

Because binary stars are so common in the Milky Way, a matter of interest in astronomy is the prevalence of planets in such systems.  Observational and theoretical studies of star formation, protoplanetary disks, and longterm dynamical stability of planets in 3-body systems have occupied the attention of many astronomers (see, e.g., the Introduction in Haghighipour 2006).

Searches for evidence of planetary systems in orbit around stars in binary systems have involved various techniques. For planets with orbital semi-major axes up to about 10 astronomical units (AU), but typically substantially less than this, precision radial velocities, transits and microlensing have been most useful (e.g., Roell et al 2012).  Techniques relevant for planets with semi-major axes of tens to hundreds of AU include direct imaging of warm planets (e.g., Vigan et al 2012; Brandt et al 2014) and of dusty debris disks (Rodriguez \& Zuckerman 2012).  However typical direct imaging campaigns have, so far, only been sensitive to planets with masses of a few Jupiters or greater and with detections too few to make any meaningful statistical comparisons between planetary systems around single stars and those in binary star systems.  Also, the existence of a dusty debris belt alone does not necessarily imply the existence of a planet.  Thus, the unearthing of any additional techniques sensitive to wide-orbit planetary systems in binary star systems would be of value. 

In the present paper we consider a new method for investigation of planets in binary star systems -- the presence or absence of heavy elements in the atmospheres of white dwarf stars.
The photospheres of at least 25\% of field white dwarfs are externally polluted with detectable quantities of elements heavier than helium (Zuckerman et al 2003, 2010; Koester et al 2014).  A wide range of evidence indicates that this pollution is from accretion of extrasolar asteroids and occasionally, perhaps, even a rocky planet (e.g., Jura 2003; Jura et al 2009; Gaensicke et al 2012; Jura \& Young 2014; Barstow et al 2014) or portion thereof (e.g., Zuckerman et al 2011). The implied mass of these extrasolar asteroid belts is often comparable to or  greater than that of the Sun's belt, while the presence of multiple major planets that gravitationally perturb the orbits of the smaller objects is likely (e.g. Zuckerman et al 2010; Debes et al 2012; Veras et al 2013, Mustill et al 2014, and references therein).  A variety of extensive theoretical calculations demonstrate that this interpretation of the observational data is sound (e.g., Rafikov 2011; Metzger et al 2012; Veras et al 2013; Mustill et al 2014; Wyatt et al 2014; Frewen \& Hansen 2014).  To the best of our knowledge, no other model for heavy element pollution of white dwarf atmospheres is now seriously considered.

One result from the Kepler satellite -- at relatively small semi-major axes, many planetary systems are densely packed -- suggests the plausibility of similarly complex systems of gravitationally interacting major planets and rocky debris at larger semi-major axes such as those sampled with white dwarf studies.  The existing data on externally polluted white dwarfs has motivated detailed models for the evolution of extended two- and three-planet systems from the main sequence through the giant branches and well into the white dwarf phase (Veras et al 2013; Mustill et al 2014).

Our primary goal is a comparison between binary stars and single stars of the frequency of planetary systems that, on the main sequence, had semi-major axes greater than a few AU.  That is, because white dwarf progenitors are A- and F-type main sequence stars, when they were red giants (on the AGB), all orbiting objects of planet mass or less with semi-major axes 
out to a few AU would have been destroyed (e.g., Jura 2008).  We consider only the situation where a planetary system orbits one member of a binary (no circumbinary planets).   Although we focus on binary star systems that contain a white dwarf and a main sequence companion, the conclusions are applicable to earlier stages of stellar evolution, specifically when the white dwarf was a main sequence star; thus, our results can be compared to studies of planets in main sequence binary star systems (see Section 4).  Note that when the progenitor of the white dwarf was on the main sequence it was the primary and the current main sequence star was the secondary.

\section{Planet frequency around single white dwarfs}

We want to compare the frequency of planetary systems around the white dwarf in a binary star system with their frequency around single white dwarfs. Planetary system occurrence frequency around single white dwarfs has been measured in the optical with spectrometers on the VLT and Keck telescopes (Zuckerman et al 2003 \& 2010) and in the ultraviolet with the COS spectrometer on the Hubble Space Telescope (Koester et al 2014).   For a wide range of white dwarf effective temperature, calcium is the element most easily detected optically, while for hot white dwarfs silicon is the element most easily detected (in the UV).

Based on results from papers cited in the third paragraph in Section 1, with a few caveats, the existence of a planetary system is considered established when either calcium or silicon is detected in the photosphere of a white dwarf.  One caveat is the possibility of radiative levitation of silicon in the atmosphere of an appropriately hot white dwarf (Chayer 2014; Koester et al 2014; Barstow et al 2014; and references therein).  Another caveat would be the  possibility of accretion by a white dwarf of interstellar rather than planetary system matter. While on occasion such a possibility cannot be conclusively ruled out, the suite of papers cited in Section 1 establishes beyond any reasonable doubt that such an occurrence would represent the exception rather than the rule.

For a sample of single DA (hydrogen atmosphere) white dwarfs with temperatures $<$10,000 K, Zuckerman at el (2003) found Ca in the photospheres of $\sim$25\%.  For a sample of single DB (helium atmosphere) white dwarfs with temperatures between 13,500 and 19,500 K, Zuckerman et al (2010) deduced that about 1/3 have Ca in their photospheres.  For a sample of DA white dwarfs with temperatures between 17,000 and 27,000 K, Koester et al (2014) found Si in the photospheres of $\sim$50\%.  Because of the possibility of radiative levitation of silicon, interpretation of this fraction is somewhat complex and we refer the reader to Koester et al (2014) and Barstow et al (2014) for details.  But, in any event, Koester et al conclude that the fraction of the COS sample with surrounding planetary systems is certainly not smaller than the fraction deduced optically in the two Zuckerman et al studies.


\section{Planet frequency around white dwarfs in binary systems}

White dwarfs in binary systems appear in a variety of classes:  well separated common proper motion (cpm) pairs, double degenerates, Sirius-like, and moderately close or very close red dwarf/white dwarf (RD/WD) pairs.  The very close RD/WD pairs and many known double degenerates have passed through a phase of common envelope evolution.  Any surviving planets in such systems will be of the circumbinary type (i.e., the planets will orbit both stars).  
For RD/WD systems the white dwarf atmosphere is often polluted with material captured from the wind of the red dwarf (see, e.g., Zuckerman et al 2003 Section 4.7).  Also, in spatially unresolved systems, Ca II K-line emission from the red dwarf oftentimes overwhelms K-line absorption in the photosphere of the white dwarf.  Because of these effects it is difficult or impossible to use spectroscopy to study planetary systems around very close RD/WD pairs.  Perhaps in some such cases planetary system material could find its way onto the white dwarf in measurable quantities (e.g., Farihi et al 2010), but these would involve careful consideration that is well beyond the scope of the present paper.

To avoid such uncertainties, we consider only white dwarf stars in cpm binary systems where the unevolved secondary star is so far ($>$120 AU, see below) from the white dwarf that there is essentially no chance that the wind of the secondary will pollute the atmosphere of the white dwarf.  Many such systems are known to exist (e.g., Table 1 in Silvestri et al 2005 and in Holberg et al 2013), but for only relatively few have the atmosphere of the white dwarf been examined for Ca with a high resolution spectrometer on a large telescope such as Keck or the VLT.  Indeed, so few (perhaps zero) white dwarfs in wide binary systems have been observed in the UV with COS that we use only Ca in our analysis described below.   Our choice of wide separation binaries pretty much guarantees that any planetary system that is discovered orbits only the white dwarf and is not circumbinary.

In Table 1 and Figure 1 we gather together data from the literature for white dwarfs in binary systems that, with but few exceptions, have been observed for the Ca II K-line with Keck and/or the VLT.  As mentioned in Sections 1 and 2, the presence of this line in the spectrum serves as a proxy for the existence of an orbiting planetary system that contains both a debris disk and at least one major planet, all with semi-major axes of at least a few AU.  The abscissa, the separation on the sky of the two stars in a given binary, is on average slightly smaller than the semi-major axis of the system.  

Given the multiple sources of data that were accessed to construct Table 1, it is not possible to control entirely against biases.  Toward this goal, inclusion of a white dwarf in the table was subject to various constraints: (1) V magnitude brighter than 17.3; (2) effective temperature (T$_{eff}$) $<$17,500 K; and (3) for cool DC and DQ white dwarfs, unambiguous indication in at least one published paper that the region of the Ca II H- and K-lines had been examined with at least a moderate resolution spectrometer.  The first two constraints relate to sensitivity to detection of the Ca K-line (more difficult for faint or hot stars).  The third constraint is to avoid inclusion of DC- and DQ-type white dwarfs that have never been observed at wavelengths shorter than 4000 \AA\ and whose atmospheres might thus actually contain calcium notwithstanding their current classifications.  (White dwarfs with atmospheric calcium include the letter Z in their classification; see the third column in Table 1).


Although one is contending with small number statistics, tentative conclusions regarding wide orbit planetary systems in binary star systems may be drawn from Figure 1.   First, for cpm binary systems composed of a white dwarf and a main sequence star with separations less than about 2500 AU, the white dwarf stars contain a smaller percentage of wide-orbit planetary systems -- 1/21 (5\%) -- than do single white dwarfs, for similar white dwarf temperature ranges.  Due to expansion of orbits during mass-losing phases of the white dwarf progenitor, current star-star separations are typically 2 to 3 times the separations when the white dwarf was on the main sequence (see, e.g., Farihi et al 2013).  Thus Figure 1 suggests that in a main sequence binary star system with semi-major axis $\leq$1000 AU and where the primary is an A- or F-type star, then extended planetary systems in orbit around the primary star are less common than around single A- and F-type stars.

Another conclusion that may be drawn from Figure 1 is that in main sequence binary systems with semi-major axes larger than about 1000 AU, the frequency of wide-orbit planetary systems around primary A- and F-type stars may well be comparable to the frequency around single stars.  That is, for the separation bins from 2500 to 9250 AU the Ca detection frequency is 5/17 (30\%), essentially the same as measured for single white dwarfs in the same temperature regime (Section 2). 

As mentioned above, our aim is to investigate the frequency of planetary systems that orbit one but not both members of a 
binary system (i.e. are not circumbinary planets).  Therefore, in construction of Table 1 and Figure 1 we avoided binary 
systems where calcium, if seen in absorption in the spectrum of the white dwarf, might be due to a circumbinary planetary system.  For sufficiently well-separated stars there is 
little chance that a planetary system will be circumbinary.  As noted just above, mass loss will cause expansion of orbits, 
typically by a factor of 2-3.  A dynamically stable circumbinary planetary system will have an orbital semi-major axis at least 
3.5 times the semi-major axis of the stellar binary (Holman \& Wiegert 1999).  Thus, any dynamically stable circumbinary 
planetary system, when on the main sequence, typically would have to have had a semi-major axis at least somewhat larger than 
the current semimajor axis of a white dwarf/main sequence binary.  The smallest apparent separation of any of the stars in Table 1 is 
the GD 165 system (120 AU separation).  Based on the semi-major axes of the few directly imaged planetary systems (e.g., Marois 
et al 2010) and of dusty debris disks in binary star systems (Section 4), only circumbinary planetary systems with 
unusually extreme (large) semi-major axes could encircle any of the systems listed in Table 1.

\section{Planet frequency in binary star systems that lack a white dwarf member}

To date the most successful planet discovery methods have been stellar 
transits and precision radial velocities (PRV) but with semi-major axes typically 
well less than an AU for the former and less than about 5 AU for the latter 
technique (e.g, Figure 1 in Roell et al 2012).  The calculations of Holman 
\& Wiegert (1999) indicate that, depending on the binary mass ratio and 
eccentricity, a planet in orbit around the primary star could be stable if 
its semi-major axis was not more than about 20\% that of the semi-major axis 
of the secondary star.  However, Figure 4 in Roell et al illustrates that 
for actual binary and triple star systems with known planets, the typical 
ratio of planet/host star separation to star-star separation is much less 
than 20\% -- more like one percent.

Typically, PRV measurements are not undertaken for binary systems with separations 
$<$2 arc sec (J. Wright 2014, private communication).   Since the distance 
to a typical star targeted with the PRV technique is 50 pc or greater, one 
anticipates projected linear separations $>$100 AU between primary and secondary star, in
agreement with the distribution of points on the abscissa in Figure 4 in Roell et al (2012).   
In Figure 2
we plot the distribution of semi-major axes of planets detected with PRV (upper line) and 
the subset that orbit one star in a binary system (lower line).  The data are from 
the website exoplanets.org.  There does not appear to be any obvious difference in 
shape between the two lines.  This suggests that -- contrary to the situation outside of a few AU that is probed by white dwarf studies -- inside of a few AU planetary orbits are 
not significantly disrupted by a companion star with semi-major axis between a few 100 and about 10000 AU (Figure 4 in Roell et al 2012).   However, some orbital characteristics that are not delineated in Figure 2, for example planetary eccentricities (e.g., Kaib et al 2013), might differ between
binary and single star systems.  Also, unlike Figure 1, Figure 2 does not delineate star-star separations.  Although well beyond the scope of the present paper, a system by system examination of those plotted in red in Figure 2 might reveal a planet occurrence frequency that is sensitive to star-star separation.

Most material accreted onto white dwarf stars is dry and rocky, 
suggestive of asteroidal material that originated inside the snow line 
(Klein et al 2011; Xu et al 2014 and references therein).  For A- and F-type stars, this 
line will lie within a few tens of AU of the star.  Based on Figure 1, 
and as noted in Section 3, heavy element pollution in white dwarfs in 
binary systems appears unlikely for star-star semi-major axes up to 
$\sim$1000 AU when the white dwarf progenitor was on the main sequence.  
The ratio of snow line radius to 1000 AU is similar to the 1\% ratio 
mentioned two paragraphs above for planets close to their host 
stars.  Simulations by Kaib et al 
(2013) in their analysis of properties of planets studied with the
PRV technique suggest that a distant stellar companion can have "dramatic" 
effects on a planetary system, strongly perturbing the system dynamically, increasing planet 
orbital eccentricities, and even causing outer planet ejection.  

Spatially resolved images of dusty debris disks around A- and F-type 
main sequence stars sometimes reveal narrow dust rings suggestive of 
shepherding planets with semi-major axes sometimes as large as $\sim$100 
AU, and even in some multiple star systems (e.g. HR 4796, Fomalhaut, and 
Figure 9 in Rodriguez and Zuckerman 2012).  It remains to be determined 
whether the dusty debris is a definite signpost for planets and, since 
stars with debris disks tend to be young, whether disappearance of the 
dust for older systems is related to the presence and characteristics of a 
secondary star.  The study of Rodriguez and Zuckerman (2012) indicated 
that debris disks are less common around stars in multiple systems than 
they are around single stars.  Herschel data will enable improved 
statistics on dusty debris in multiple star systems (D. Rodriguez et al, 
in preparation).

To date roughly 50 planets have been discovered through microlensing, a 
few of which are in binary systems (e.g., Gould et al 2014).  
However, because of certain biases in the first generation surveys, 
binaries were discriminated against.  Revisions in the observational 
procedures should eliminate these biases and, in the near future, 
microlensing should become a good method for detection of planets in 
binary star systems (A. Gould, private communication).  For a number of 
reasons, microlensing is more sensitive to planets that orbit late G- 
through M-type stars than to the earlier type stars that are probed by 
studies of externally polluted white dwarfs.  And microlensing has 
good sensitivity to 
planets with semi-major axes in the vicinity of the snow line, so it and 
the white dwarf technique should complement each other.

\section{CONCLUSIONS}

The occurrence of heavy elements, most notably calcium or silicon, in 
the atmospheres of white dwarf stars can be used to determine the 
relative frequency of wide-orbit ($>$few AU) planets around main sequence single 
stars and members of binary systems. We used Ca II K-line data from the 
published 
literature to conclude that, for main sequence binaries with A- and 
F-type primaries and with semi-major axes $\leq$1000 AU, the presence of 
a second star appears to suppress the formation and/or long term orbital 
stability of wide-orbit planets around the primary star 
-- even planets that have semi-major axes much smaller than that of the 
secondary.  For main sequence binaries with separations $\geq$1000 AU 
the presence of a second star does not seem to diminish the likelihood 
of wide-orbit planetary systems.

The published data employed to construct Figure 1 were not collected 
with a view toward understanding the frequency of binary star planets.  
Thus, the statistics presented here can be much improved by a dedicated 
spectroscopic campaign to measure, with large ground-based telescopes and with HST, 
white dwarfs in binary systems.  

$\smallskip$

I thank Drs. D. Koester, A. Kawka, J. Wright, and Ms. L. Vican for 
their generous assistance, Drs. B. Klein, A. Gould, G. Marcy, and S. Xu for helpful 
advice, and the referee for suggestions that improved the paper.  This 
research was supported by NASA grants to UCLA.

\section*{REFERENCES}
\begin{harvard}

\item[Aannestad, P. \& Sion, E. 1985, AJ 90, 1832] 
\item[Barstow, M., Barstow, J., Casewell, S., Holberg, J. \& Hubeny, I. 2014, MNRAS 440, 1607]
\item[Brandt, T., Kuzuhara, M., McElwain, M. et al 2014, ApJ 786, 1]
\item[Chayer,P. 2014, MNRAS 437, L95]
\item[Debes, J., Walsh, K. \& Stark, C. 2012, ApJ 747, 148]
\item[Dufour, P., Bergeron, P., Liebert, J. et al 2007, ApJ 663, 1291]
\item[Farihi, J., Bond, H. Dufour, P. et al 2013, MNRAS 430, 652]
\item[Farihi, J., Burleigh, M., Holberg,, J., Casewell, S. \& Barstow, M. 2011, MNRAS 417, 1735] 
\item[Farihi, J., Hoard, D. \& Wachter, S. 2010, ApJS 190, 275]
\item[Frewen, S. \& Hansen, B. 2014, MNRAS 439, 2442]
\item[Gaensicke, B., Koester, D., Farihi, J. et al 2012, MNRAS 424,333]
\item[Gould, A., Udalski, A., Shin, I.-G. et al 2014, Science 345, 46]
\item[Haghighipour, N. 2006, ApJ 644, 543]
\item[Holberg, J., Oswalt, T., Sion, E., Barstow, M. \& Burleigh, M. 2013, MNRAS 435, 2077]
\item[Holman, M. \& Wiegert, P. 1999, AJ 117, 621]
\item[Jura, M. 2003, ApJ 584, L91]
\item[Jura, M. 2008, AJ 135, 1735]
\item[Jura, M. Muno, M., Farihi, J. \& Zuckerman, B. 2009, ApJ 699, 1473]
\item[Jura, M. \& Young, E. 2014, Ann. Rev Earth Planet Sci, 42, in press]
\item[Kaib, N., Raymond, S. \& Duncan, M. 2013, Nature 493, 381]
\item[Kawka, A. \& Vennes, S. 2012, MNRAS 425, 1394]
\item[Klein, B., Jura, M., Koester, D. \& Zuckerman, B. 2011, ApJ 741, 64]
\item[Koester, D., Gaensicke, B. \& Farihi, J. 2014, arXiv1404.2617]
\item[Koester, D., Voss, B., Napiwotzki, R. et al 2009, A\&A 505, 441]
\item[Marois, C., Zuckerman, B., Konopacky, Q., Macintosh, B. \& Barman, T. 2010, Nature 468, 1080]
\item[Metzger, B., Rafikov, R. \& Bochkarev, K. 2012, MNRAS 423, 505]
\item[Mustill, A., Veras, D. \& Villaver, E. 2014, MNRAS 437, 1404]
\item[Rafikov, R. 2011, MNRAS 416, L55]
\item[Rodriguez, D. \& Zuckerman, B. 2012, ApJ 745, 147]
\item[Roell, T., Neuhauser, R., Seifahrt, A. \& Mugrauer 2012, A\&A 542,  A92]
\item[Silvestri, N., Hawley, S. \& Oswalt, T. 2005, AJ 129, 2428]
\item[Subasavage, J., Henry, T., Bergeron, P. et al 2007, AJ 134, 252]
\item[Subasavage, J., Henry, T., Bergeron, P. et al 2008, AJ 136, 899]
\item[Veras, D., Mustill, A., Bonsor, A. \& Wyatt, M. 2013 MNRAS 431, 1686]
\item[Vigan, A., Patience, J., Marois, C. et al 2012, A\&A 544, A9]
\item[Voss, B., Koester, D., Napiwotzki, R., Christlieb, N. \& Reimers, D. 2007, A\&A 470, 1079]
\item[Wyatt, M. Farihi, J., Pringle, J. \& Bonsor, A. 2014, MNRAS 439, 3371]
\item[Xu, S., Jura, M., Koester, D., Klein, B. \& Zuckerman, B. 2014, ApJ 783, 79]
\item[Zuckerman, B., Koester, D., Dufour, P. et al 2011 ApJ 739, 101]  
\item[Zuckerman, B., Koester, D., Reid, I. N. \& Hunsch, M. 2003, ApJ 596, 477]
\item[Zuckerman, B., Melis, C., Klein, B., Koester, D. \& Jura, M. 2010, ApJ 722, 725]

\end{harvard}

\clearpage

\begin{table}
\caption{White dwarfs in common proper motion binary systems}
\begin{tabular}{@{}lcccccccc}
\br
WD& name& type & V & T$_{eff}$& Ca EW& [Ca/H(e)]& separation& ref\\
 &  & &  (mag) & (K)&  (m\AA) &  & arcsec  AU&  \\
\mr
0119-004 & LP 587-44 & DB & 16.3 & 16540 & $<$60  &  $<$-8.8  & 9 \ \ \ \  990 &  7,8  \\
0120-024 & NLTT 4615 & DA & 17.0 & 5840 & $<$450 & $<$-10.0 & 44\ \ \ \ 1850 & 9,11 \\
0148+641& G244-36 & DA& 14.0 &  6100(?)&  $<$5&	$<$-12.1 & 12 \ \ \ \  200 & 1\\
0204-306 &  NLTT 7051 & DA  & 16.2 & 5640 & $<$500 & $<$-10.0 & 73 \ \ \ \  2190 &  9,11 \\
0250-007&  LP 591-177 &	DA & 16.4 & 8400 &  $<$90  &	$<$-9.6	&	27\ \ \ \ 1020 & 	5,7 \\
0415-594 & $\eta$ Ret B & DA & 12.5 &  15310 & & $<$-9.0	  &	12.8\ \ \ \ 240 &	6\\
0433+270 & G39-27 & DA & 15.6 & 5430(?) & $<$30  &      $<$-12.2  & 124	\ \ \ \  2230 & 1 \\
0615-591 &  L182-61 &	DB & 14.0  & 16710 & $<$30  & $<$-9.4 &  41\ \ \ \	1540	  & 7,8 \\
0625+100 & G105-B2B & DZ & 16.5 &   & & & 127 \ \ \ \ 7880 & 2 \\
0642-285	& LP895-41 &	DA & 15.2 &  9370 & $<$70  &  $<$-9.3	&  16\ \ \ \ 	510 &	5,7 \\
0738-172 & LHS 235 & DZA & 13.0 & 7650 & 6000 & -10.9 & 21.4 \ \ \ \  200 & 13 \\
0751-252 &  SCR0753-2524 & DC & 16.3 & 5080 &  &  &  396 \ \ \ \  7740 &  14 \\
0845-188 & LP786-6 & DB & 15.7 & 17450 & $<$15 &	$<$-9.2  &	 31\ \ \ \ 8740 &	3\\
1004+665	& LP 62-35 &	DZ	& 14.7 & 	&	&	& 112\ \ \ \ 7050  &  10 \\
1009-184	& LHS 2033B &	 DZ & 15.4  & 9940 &   &  &	400\ \ \ \	6870 &  4 \\
1105-048 & NLTT 26379 &  DA  & 13.1 & 16110 & $<$10  & $<$-8.5 & 279\ \ \ \ 7200 & 5,7 \\
1147+255	& G121-22 & 	DA & 15.6 &  10260 & $<$15 & $<$-9.9 & 36\ \ \ \     1980 &	1\\
1209-060 & LP 674-29 & DA & 17.3 & 6400 & $<$60 & $<$-10.3 & 203\ \ \ \ 9170 & 9,11 \\
1327-083 &  LHS 354 & DA  & 12.3 & 14570 & $<$7  & $<$-8.9  &  503\ \ \ \ 9070 & 5,7 \\
1336+123 & LP 498-26 & DB & 14.8 & 16780 & $<$25 & $<$-9.5 & 87 \ \ \ \  4440 & 7,8 \\
1345+238	& LHS 361 &	DA & 15.6 &  4700 & $<$40 & $<$-12.8 &	198\ \ \ \  2400 & 1\\
1348-272 & LP 856-53 & DA & 14.5 & 9830 & $<$40 & $<$-9.6  & 9 \ \ \ \  280 & 5,7 \\
1422+095 & GD165 & DAV & 14.4 & 12440 & $<$20  &  $<$-8.8 & 3.7\ \ \ \ 120 & 1,5\\
1425+540	& G200-040 &	DBAZ & 15.0 & 14750 &  150 &	-9.3	& 60\ \ \ \   3480 &	3\\
1542-275 &  LP 916-27 & DB & 15.4 & 10800 & $<$30 & $<$-11.8  & 54\ \ \ \ 2800 & 7,8  \\
1544-377	& LTT 6302 &	DA &  12.8 & 11270 &	$<$3 & $<$-10.3 & 15\ \ \ \ 240	 &	1\\
1555-089  & G152-B4A & DA & 14.7 & 14530 & $<$30 &  $<$-8.1 & 11 \ \ \ \ 560 &  5,7 \\
1619+123	& PG &	DA & 14.7 &  17150 & $<$30	& $<$-7.5 & 62\ \ \ \    3220 &	5,7 \\
1716+020	& Wolf 672 & DA & 14.3 &  13620 & $<$20 & $<$-8.5 & 13\ \ \ \ 460  &	1 \\
1911+135 & G142-B2A & DA & 14.2 & 13780 & $<$30 & $<$-8.3 & 19 \ \ \ \ 630 & 5,7 \\
1917-077 & LTT 7658 & DBQA & 12.3 & 10200 & $<$15 & $<$-12.1 & 27\ \ \ \ 310 & 7,8 \\
1932-136 & L852-37 & DA & 15.9 & 16930 & $<$30 & $<$-7.4 & 29 \ \ \ \ 3080 & 5,7 \\
2051+095 & LP 516-13 & DA & 16.0 & 15670 & $<$30 &$<$-8.1  & 10 \ \ \ \ 1030 & 5,7 \\
2129+000	& G26-10	& DB & 14.7 &  14000 &  $<$7 &	$<$-11.1 &  133\ \ \ \   6020 &	3\\
2253-081	& NLTT 55288	& DA  & 16.5 & 6770 &  $<$100 &  $<$-10.0 &	42\ \ \ \  1540 &	5,7 \\
2253+803 & HS2253+8023 & DBAZ & 16.1 & 14400 & 4700 & -7.0 & 39 \ \ \ \  2730  & 12 \\
2318+126 & LP 522-34 & DA & 15.9 & 14020 & $<$30 & $<$-8.7 & 35 \ \ \ \  3050 &  5,7 \\
2341+322	& LP 347-4 &	DA  & 12.9 & 12577& $<$5	& $<$-9.5 &	175\ \ \ \  3080 &	1 \\
\br
\end{tabular}
\end{table}

\clearpage

\noindent Notes to Table 1 $-$ Ca EW refers to the equivalent width of the Ca II K-line. [Ca/H(e)] is the logarithm of the ratio by number of Ca atoms to either H or He atoms depending on which is the dominant constituent of the white dwarf atmosphere. 1 = Zuckerman et al 2003; 2 = Aannestad \& Sion 1985; 3 = Zuckerman et al 2010;
4 = Subasavage et al 2007; 5 = Koester et al 2009; 6 = Farihi et al 2011; 7 = D. Koester 2014 (private communication); 8 =  Voss et al 2007; 9 = Kawka \& Vennes 2012;  10 = Holberg et al 2013; 11 = A. Kawka 2014 (private communication); 12 = Klein et al 2011; 13 = Dufour et al 2007; 14 = Subasavage et al 2008

\clearpage

\begin{figure}
\includegraphics[width=140mm]{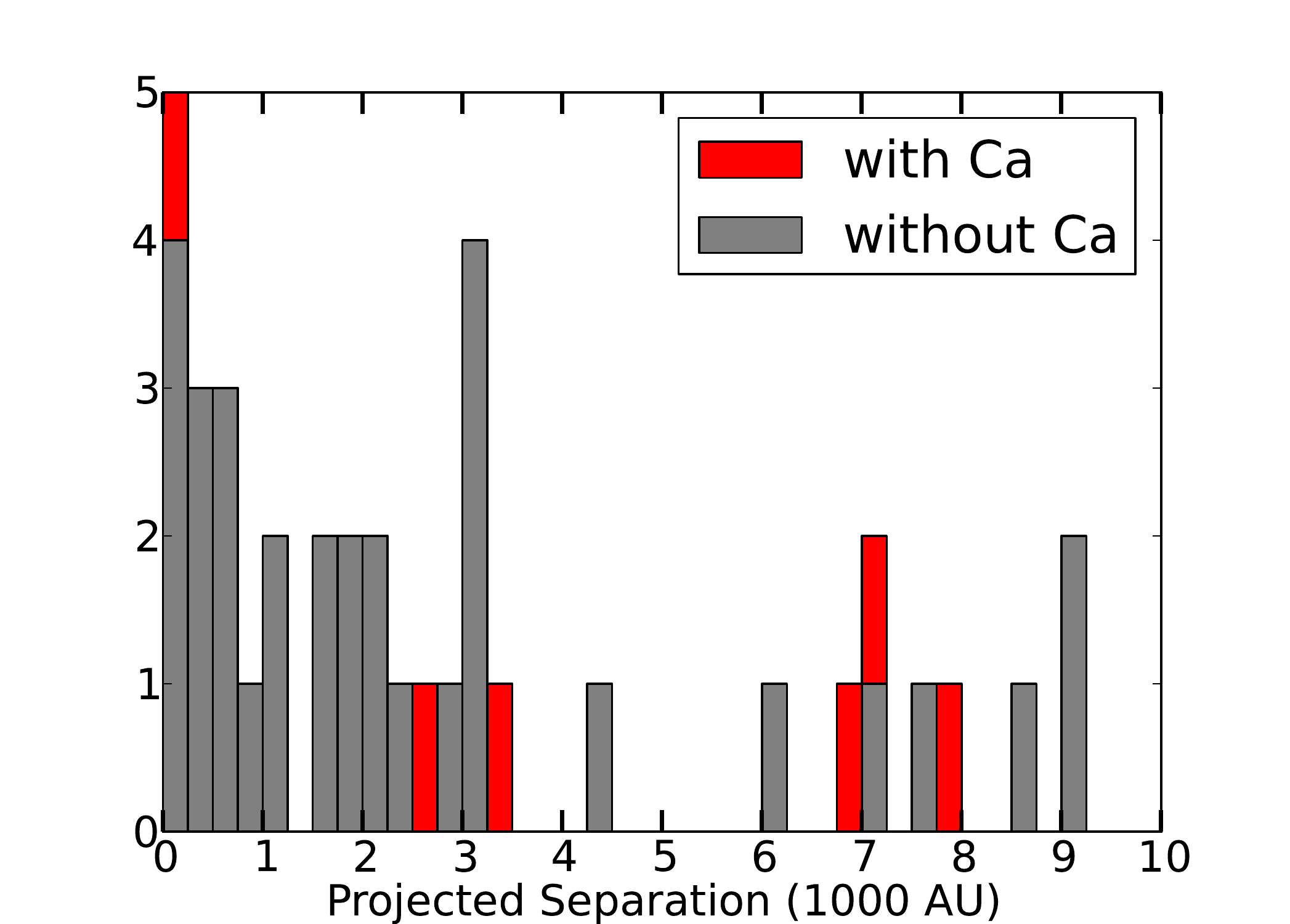}
\caption{\label{figure1} White dwarf stars in binary systems with (red) and without (gray) detected photospheric Ca II K-line absorption.   The abscissa is the separation in the plane of the sky between the white dwarf and its main sequence companion.  The ordinate is the number of white dwarfs in each of the separation bins of width 250 AU, but where systems with separations $<$120 AU have been excluded (Section 3).}  
\end{figure}

\clearpage

\begin{figure}
\includegraphics[width=140mm]{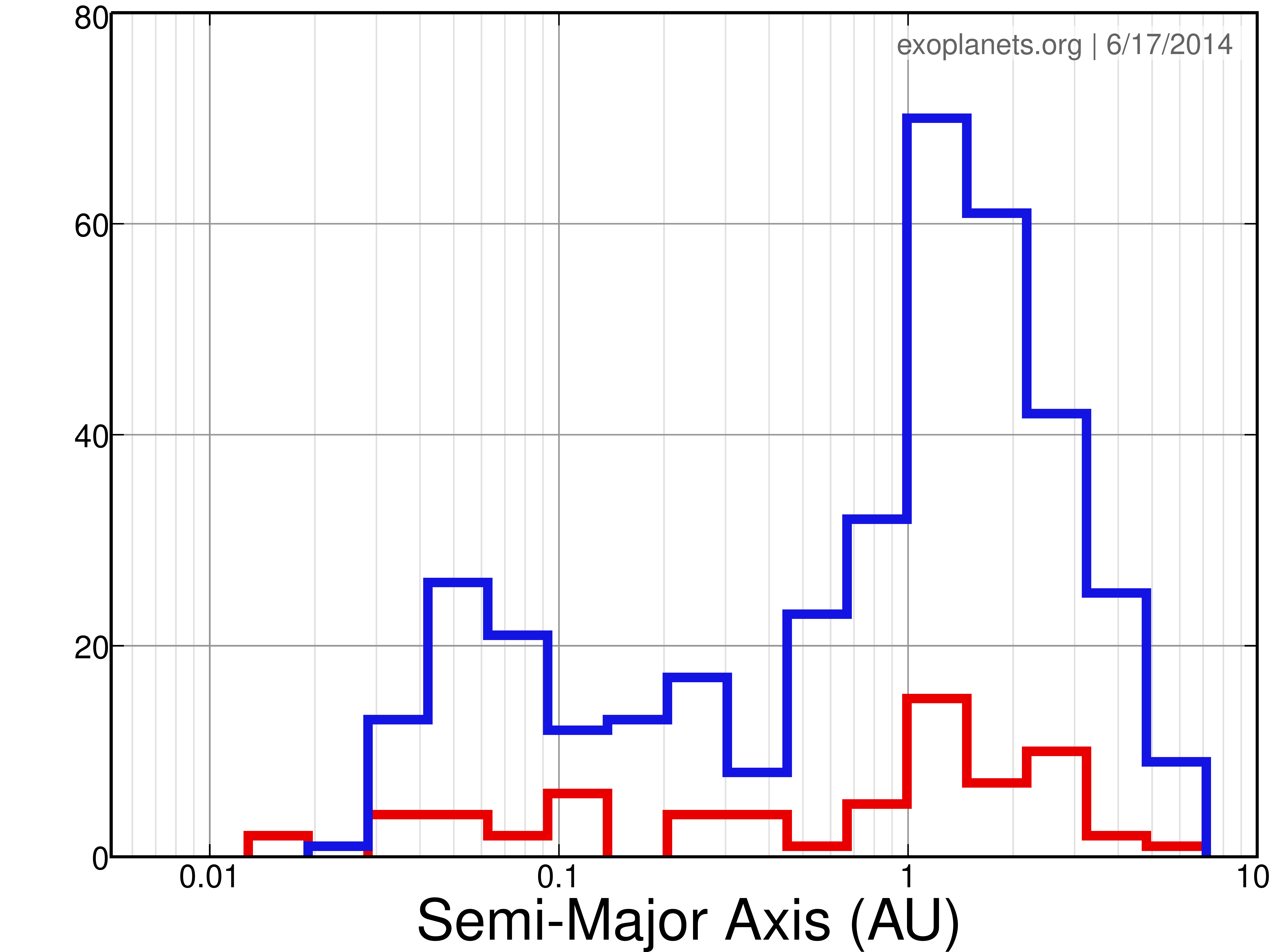}
\caption{\label{figure2} Histogram of number of planets detected with the precision radial velocity (PRV) technique vs planet semimajor axis.  The upper (blue) line is the planet distribution for all PRV detected planets and the lower (red) line is for those in binary star systems with star-star semi-major axes that fall in the same range as in Figure 1 (see Figure 4 in Roell et al 2012).  At the small semi-major axes shown here the stellar secondary does not appear to have a noticeable effect on the distribution of semi-major axes of the detected planets.}  
\end{figure}

\end{document}